\documentclass{aa}
\usepackage{graphics}
\begin{document}

%
%
\title{The emission spectrum of the strong Fe\,II emitter BAL Seyfert 1 galaxy 
IRAS\,07598+6508}

\author{ M.-P. V\'eron-Cetty \inst{1}, M. Joly \inst{2}, P. V\'eron \inst{1}, T. 
Boroson \inst{3}, S. Lipari \inst{4} and P. Ogle \inst{5}}  

\institute
{Observatoire de Haute Provence, CNRS, F-04870 Saint-Michel l'Observatoire,
    France\\ 
\email{mira.veron@oamp.fr; philippe.veron@oamp.fr}
\and
Observatoire de Paris-Meudon, 5 place J. Janssen, F-92195 Meudon, France\\
\email{Monique.Joly@obspm.fr}
\and
National Optical Astronomy Observatory, 950 North Cherry Avenue, Tucson, AZ 85719\\
\email{tyb@noao.edu}
\and
Cordoba observatory and CONICET, Laprida 854, 5000 Cordoba, Argentina\\
\email{lipari@mail.oac.uncor.edu}
\and
Spitzer Science Center, California Institute of Technology, MS 220-6, Pasadena, CA 91125\\
\email{ogle@ipac.caltech.edu}}

\titlerunning{IRAS\,07598+6508}
\authorrunning{V\'eron-Cetty et al.}

\date{Received ; accepted }

\abstract{The narrow-line Seyfert 1 galaxy IRAS\,07598+6508 is known to be a  
stong Fe\,II emitter. The analysis of several high S/N ratio spectra shows that 
its spectrum is dominated by a relatively narrow "broad line" region
(1\,780 km s$^{-1}$ FWHM) emitting not only Fe\,II, but also Ti\,II and Cr\,II 
lines. Although we were unable to find a completely satisfactory physical model, 
we got the best agreement with the observations with collisional rather than 
radiative models, with a high density (n=10$^{15}$ cm$^{-3}$), a high column 
density (N$_{H}$=10$^{25}$ cm$^{-2}$) and a microturbulence of 100 km s$^{-1}$. 
This BLR is qualitatively similar to the one observed in I\,Zw\,1. We have not 
found traces in IRAS\,07598+6508 of the narrow line regions found in I\,Zw\,1.
}

\maketitle
\today

\section{Introduction}

 IRAS\,07598+6508 was identified as a 14.3 mag. starlike AGN candidate 
by de Grijp et al. (\cite{grijp87}). Optical and infrared images are dominated by the 
point-source nucleus (Scoville et al. \cite{scoville00}; Surace \& Sanders 
\cite{surace00}; Kim et al. \cite{kim02}). An HST image, published by Boyce 
et al. (\cite{boyce96}) and Canalizo \& Stockton (\cite{canalizo00}), shows a great 
number of knots, presumably OB associations. 

 This object was shown to be a Seyfert 1 galaxy by Sanders et al.
(\cite{sanders88}). Its redshift, z=0.149, corresponds to a distance of 630 Mpc {\footnote{Throughout this paper, 
we use H$_{\rm o}$=70 km s$^{-1}$ Mpc$^{-1}$}. It is a very strong Fe\,II emitter 
(Lawrence et al. \cite{lawrence88}; Low et al. \cite{low88}; \cite{low89}) and a BAL 
QSO (Boroson \& Meyers \cite{boroson92}). 
It is an ultraluminous IR galaxy with a bolometric luminosity greater than 
10$^{12}$ L$_\odot$ (Sanders et al. \cite{sanders88}). It is a weak hard X-ray 
source (Gallagher et al. \cite{gallagher99}; Green et al. \cite{green01}; Imanishi 
\& Terashima \cite{imanishi04}). It contains a weak compact ($<$0$\farcs$1) nuclear
radio source (S$_{15 GHz}$=2.9 mJy)(Nagar et al. \cite{nagar03}). 

 The aims of this paper are to identify all emission lines in its spectrum 
and to try to determine the physical conditions in the emission line regions.  
Similarities and differences with I\,Zw\,1, another well studied object, will also be noted.

\section{The data}

\subsection{The Keck 10-m telescope spectrum}

 A 2400 s spectrum was obtained on January 28, 1995 with the Keck 10-m 
telescope on Mauna Kea and the Low Resolution Imaging Spectrograph (LRIS, Oke 
et al. \cite{oke95}) and polarimeter combination (Ogle et al. \cite{ogle99}). 
A 300 groove mm$^{-1}$ grating with a dispersion of 2.49 \AA\ pixel$^{-1}$ 
gave a resolution of 10 \AA. The spectrum covers the observed wavelength range 
3800-8900 \AA. The detector was a 2048$\times$2048 24 $\mu$m pixel Tektronix 
CCD.


\begin{table}[ht]
\caption{\label {spectra} Available spectra of IRAS\,07598+6508. Col. 1: telescope, 
col. 2: observed spectral range, col. 3: exposure time, col. 4: resolution (FWHM \AA),
col. 5: S/N ratio in the range 5100-5300 \AA.}
\begin{center}
\begin{tabular}{|l|c|c|c|r|}
\hline
 Telescope & spec. range & exp. t. (s) & res. (\AA) & S/N \\
\hline
KPNO 4-m   & 5590-8220 & 2700 &   3 & 100 \\
KPNO 2.1-m & 3900-5890 & 3000 &   7 & 100 \\
KPNO 2.1-m & 5830-7810 & 2400 &   7 & 100 \\
Keck 10-m  & 3800-8900 & 2400 &  10 & 180 \\
Bok 2.3-m  & 4500-8000 & 6400 &  12 & 600 \\
\hline
\end{tabular}
\end{center}
\end{table}

\subsection{The KPNO 4-m telescope spectrum}

 A 2\,700 s spectrum was obtained on February 1, 1991 with the KPNO 4-m 
telescope and the RC 
spectrograph (Boroson \& Meyers \cite{boroson92}). The detector was a
thick Tektronics 2048$\times$2048 21\,$\mu$m pixel CCD. A 600 grooves 
mm$^{-1}$ grating was used in first order, giving a dispersion of about 1.3 
\AA\ pixel$^{-1}$. The spectral coverage was 2\,630 \AA\ (5590-8220 \AA). A 
slit width of 225 $\mu$m, corresponding to about 1$\farcs$5, was used. This 
projected on the CCD to a  FWHM of 2.3 pixels or 3.0 \AA\ as measured from 
comparison lines. Spectrophotometric standards were observed. These star 
observations were used for removal of atmospheric features as well as flux 
calibration.

\begin{figure}[h]

\resizebox{8.8cm}{!}{\includegraphics{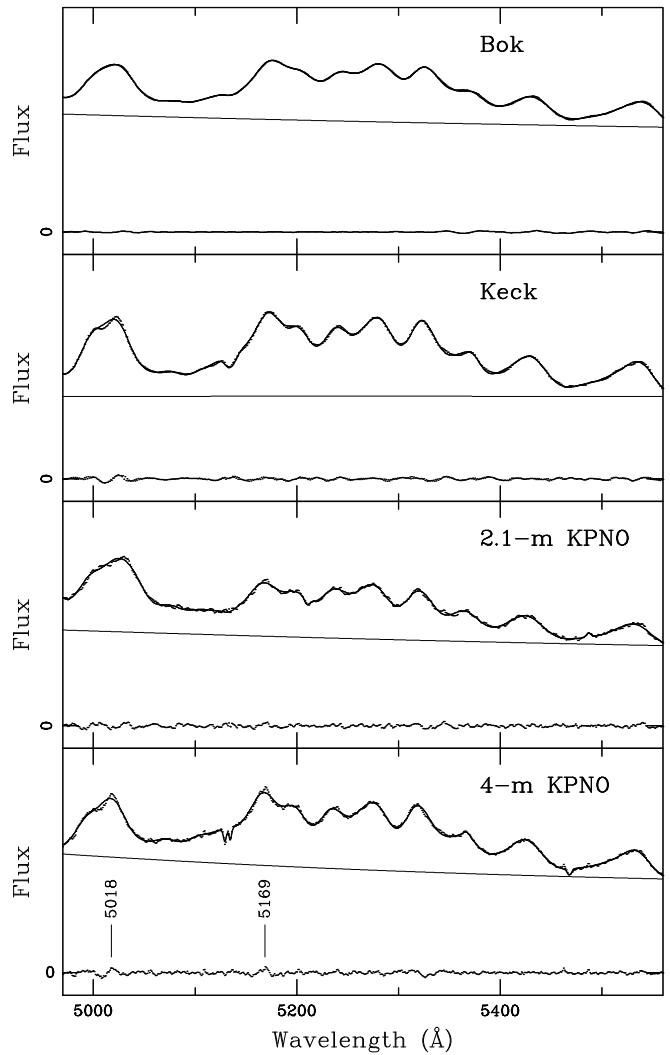}}

\caption{\label{SP_IRAS} The deredshifted spectra of IRAS\,07598+6508 in the range
$\lambda\lambda$4970-5570, with the fits, the continua and the residuals. The positions 
of the Fe\,II 42 $\lambda$5018 and $\lambda$5169 lines are shown. The residuals of 
the 4-m KPNO spectrum at these positions are large, probably indicating that the 
profile of these lines is not trully Gaussian (see text).}

\end{figure}

\subsection{The 2.3-m Bok telescope spectrum}

 A 6\,400 s spectrum was obtained on November 18, 1995 with the 2.3-m Bok 
telescope on Kitt Peak and the spectropolarimeter described by Schmidt et al. 
(\cite{schmidt92}). It has been published by Schmidt \& Hines (\cite{schmidt99}).
The detector was a 800$\times$1200 15 $\mu$m pixel Loral CCD. A 600 grooves 
mm$^{-1}$ grating was used. A spectral coverage of about 4\,500-8\,000 \AA\ was 
obtained. A slit width of 3$\arcsec$ was used. The spectral resolution was 
$\sim$12 \AA. Spectrophotometric standards were observed. These star observations 
were used for flux calibration. 

\subsection{The KPNO 2.1-m telescope spectra}

 Two spectra were obtained on February 15, 1991 with the KPNO 2.1-m telescope and 
the Gold spectrograph. The detector was a TI 800$\times$800 15 $\mu$m pixel CCD. The 
observations were made using a 300 grooves mm$^{-1}$ grating. The blue spectrum was 
exposed for 2$\times$1\,500 s. It covers the range 3900-5890 \AA. The red spectrum was 
exposed for 2$\times$1\,200 s and covers the range 5830-7810 \AA. The resolution was 
6.5-7.0 \AA. These spectra were published by Lipari et al. (\cite{lipari93}) and 
Lipari (\cite{lipari94}).


\begin{table}[ht]
\caption{\label {tfe2} Observed permitted Fe\,II multiplets in the spectrum of 
IRAS\,07598+6508. Col. 1: multiplet number, col. 2: transition, col. 3: upper 
level energy, col. 4: number of observed lines/number of lines in the multiplet
in the observed spectral range.}
\begin{center}
\begin{tabular}{|r|c|c|c|}
\hline
 m. & Transition & u.l.(eV) &  \\
\hline
   3 & a$^{4}$P-z$^{6}$D$^{\rm o}$ & 4.80 & 1/8  \\
  11 & a$^{2}$P-z$^{6}$D$^{\rm o}$ & 4.80 & 1/1  \\
  24 & b$^{4}$P-z$^{6}$D$^{\rm o}$ & 4.80 & 1/8  \\
  34 & b$^{4}$F-z$^{6}$D$^{\rm o}$ & 4.80 & 3/4  \\
  40 & a$^{6}$S-z$^{6}$D$^{\rm o}$ & 4.80 & 3/3  \\
\hline
  12 & a$^{2}$P-z$^{6}$F$^{\rm o}$ & 5.23 & 1/5  \\
  17 & a$^{2}$H-z$^{6}$F$^{\rm o}$ & 5.23 & 1/5  \\
  25 & b$^{4}$P-z$^{6}$F$^{\rm o}$ & 5.23 & 6/8  \\
  35 & b$^{4}$F-z$^{6}$F$^{\rm o}$ & 5.23 & 9/12 \\
  41 & a$^{6}$S-z$^{6}$F$^{\rm o}$ & 5.23 & 2/3  \\
  46 & a$^{4}$G-z$^{6}$F$^{\rm o}$ & 5.23 & 6/11 \\
\hline
  26 & b$^{4}$P-z$^{6}$P$^{\rm o}$ & 5.34 & 4/6  \\
  31 & a$^{4}$H-z$^{6}$P$^{\rm o}$ & 5.34 & 1/3  \\
  36 & b$^{4}$F-z$^{6}$P$^{\rm o}$ & 5.34 & 5/5  \\
  42 & a$^{6}$S-z$^{6}$P$^{\rm o}$ & 5.34 & 3/3  \\
  47 & a$^{4}$G-z$^{6}$P$^{\rm o}$ & 5.34 & 1/3  \\
\hline
  21 & a$^{2}$D-z$^{4}$D$^{\rm o}$ & 5.56 & 4/6  \\
  27 & b$^{4}$P-z$^{4}$D$^{\rm o}$ & 5.56 & 8/8  \\
  33 & a$^{4}$H-z$^{4}$D$^{\rm o}$ & 5.56 & 1/3  \\
  38 & b$^{4}$F-z$^{4}$D$^{\rm o}$ & 5.56 & 8/9  \\
  43 & a$^{6}$S-z$^{4}$D$^{\rm o}$ & 5.56 & 2/3  \\
  48 & a$^{4}$G-z$^{4}$D$^{\rm o}$ & 5.56 & 5/6  \\
  56 & b$^{2}$H-z$^{4}$D$^{\rm o}$ & 5.56 & 1/1  \\
\hline
  28 & b$^{4}$P-z$^{4}$F$^{\rm o}$ & 5.57 & 6/6  \\
  32 & a$^{4}$H-z$^{4}$F$^{\rm o}$ & 5.57 & 5/6  \\
  37 & b$^{4}$F-z$^{4}$F$^{\rm o}$ & 5.57 & 10/10 \\
  44 & a$^{6}$S-z$^{4}$F$^{\rm o}$ & 5.57 & 1/1  \\
  49 & a$^{4}$G-z$^{4}$F$^{\rm o}$ & 5.57 & 9/9  \\
  55 & b$^{2}$H-z$^{4}$F$^{\rm o}$ & 5.57 & 2/3  \\
\hline
  16 & a$^{2}$P-z$^{4}$P$^{\rm o}$ & 5.85 & 2/3  \\
  23 & a$^{2}$D-z$^{4}$P$^{\rm o}$ & 5.85 & 1/5  \\
  29 & b$^{2}$P-z$^{4}$P$^{\rm o}$ & 5.85 & 7/7  \\
  74 & b$^{4}$D-z$^{4}$P$^{\rm o}$ & 5.85 & 8/8  \\
\hline
 148 & c$^{2}$D-z$^{2}$D$^{\rm o}$ & 7.54 & 1/4  \\
\hline
 114 & c$^{2}$G-z$^{2}$G$^{\rm o}$ & 7.66 & 2/4  \\
\hline
 182 & d$^{2}$D-z$^{2}$P$^{\rm o}$ & 8.00 & 1/3  \\
\hline
\end{tabular}
\end{center}
\end{table}


\begin{table}[ht]
\caption{\label {tti2} Observed permitted Ti\,II multiplets in the spectrum of 
IRAS\,07598+6508. Col. 1: multiplet number, col. 2: transition, col. 3: upper 
level energy, col. 4: number of observed lines/number of lines in the multiplet
in the observed spectral range.}
\begin{center}
\begin{tabular}{|r|c|c|c|}
\hline
 m. & Transition & u.l.(eV) &   \\
\hline
   6 & b$^{4}$F-z$^{4}$G$^{\rm o}$ & 3.67 & 2/8  \\
  11 & a$^{2}$F-z$^{4}$G$^{\rm o}$ & 3.67 & 3/5  \\
  17 & a$^{2}$D-z$^{4}$G$^{\rm o}$ & 3.67 & 1/3  \\
     & a$^{4}$P-z$^{4}$G$^{\rm o}$ & 3.67 & 1/3  \\
\hline
  12 & a$^{2}$F-z$^{4}$F$^{\rm o}$ & 3.84 & 2/6  \\
  18 & a$^{2}$D-z$^{4}$F$^{\rm o}$ & 3.84 & 2/5  \\
  68 & b$^{2}$D-z$^{4}$F$^{\rm o}$ & 3.84 & 2/5  \\
  80 & a$^{2}$H-z$^{4}$F$^{\rm o}$ & 3.84 & 2/3  \\
\hline
  13 & a$^{2}$F-z$^{2}$F$^{\rm o}$ & 3.87 & 2/4  \\
  19 & a$^{2}$D-z$^{2}$F$^{\rm o}$ & 3.87 & 3/3  \\
  31 & a$^{2}$G-z$^{2}$F$^{\rm o}$ & 3.87 & 3/3  \\
  69 & b$^{2}$D-z$^{2}$F$^{\rm o}$ & 3.87 & 1/3  \\
\hline
  14 & a$^{2}$F-z$^{2}$D$^{\rm o}$ & 3.94 & 1/2  \\
  20 & a$^{2}$D-z$^{2}$D$^{\rm o}$ & 3.94 & 4/4  \\
  40 & a$^{4}$P-z$^{2}$D$^{\rm o}$ & 3.94 & 4/5  \\
  50 & a$^{2}$P-z$^{2}$D$^{\rm o}$ & 3.94 & 2/3  \\
  70 & b$^{2}$D-z$^{2}$D$^{\rm o}$ & 3.94 & 1/4  \\
\hline
  15 & a$^{2}$F-z$^{4}$D$^{\rm o}$ & 4.04 & 5/5  \\
  21 & a$^{2}$D-z$^{4}$D$^{\rm o}$ & 4.04 & 3/6  \\
  41 & a$^{4}$P-z$^{4}$D$^{\rm o}$ & 4.04 & 7/8  \\
  51 & a$^{2}$P-z$^{4}$D$^{\rm o}$ & 4.04 & 3/5  \\
  61 & b$^{4}$P-z$^{4}$D$^{\rm o}$ & 4.04 & 5/8  \\
\hline
  34 & a$^{2}$G-z$^{2}$G$^{\rm o}$ & 4.26 & 1/4  \\
  82 & a$^{2}$H-z$^{2}$G$^{\rm o}$ & 4.26 & 2/3  \\
\hline
  52 & a$^{2}$P-z$^{2}$S$^{\rm o}$ & 4.64 & 1/2  \\
  92 & b$^{2}$P-z$^{2}$S$^{\rm o}$ & 4.64 & 1/2  \\
\hline
  72 & b$^{2}$D-y$^{2}$D$^{\rm o}$ & 4.85 & 2/4  \\
  93 & b$^{2}$P-y$^{2}$D$^{\rm o}$ & 4.85 & 2/3  \\
\hline
  94 & b$^{2}$P-z$^{2}$P$^{\rm o}$ & 4.89 & 3/4  \\
\hline
  75 & b$^{2}$D-y$^{2}$F$^{\rm o}$ & 4.94 & 2/3  \\
  87 & b$^{2}$G-y$^{2}$F$^{\rm o}$ & 4.94 & 2/3  \\
\hline
  76 & b$^{2}$D-y$^{4}$D$^{\rm o}$ & 5.04 & 3/6  \\
\hline
 104 & b$^{2}$F-y$^{2}$G$^{\rm o}$ & 5.42 & 2/3  \\
\hline
 105 & b$^{2}$F-x$^{2}$D$^{\rm o}$ & 5.57 & 3/3  \\
 113 & c$^{2}$D-x$^{2}$D$^{\rm o}$ & 5.57 & 1/4  \\
\hline
  99 & b$^{2}$P-y$^{2}$P$^{\rm o}$ & 5.62 & 1/3  \\
 114 & c$^{2}$D-y$^{2}$P$^{\rm o}$ & 5.62 & 2/3  \\
\hline
 115 & c$^{2}$D-x$^{2}$F$^{\rm o}$ & 5.42 & 2/3  \\
\hline
\end{tabular}
\end{center}
\end{table}

\begin{table}[ht]
\caption{\label{tcr2} Observed permitted Cr\,II multiplets in the spectrum of 
IRAS\,07598+6508. Col. 1: multiplet number, col. 2: transition, col. 3: 
upper level energy, col. 4: number of observed lines/number of lines in the multiplet.}
\begin{center}
\begin{tabular}{|r|c|c|c|}
\hline
 m. & Transition & u.l.(eV) & \\
\hline
   1 & a$^{4}$D-z$^{6}$F$^{\rm o}$ & 5.86 & 1/11 \\
\hline
   2 & a$^{4}$D-z$^{6}$P$^{\rm o}$ & 6.00 & 1/8  \\
\hline
  12 & a$^{4}$P-z$^{4}$P$^{\rm o}$ & 6.09 & 7/7  \\
\hline
  19 & b$^{4}$D-z$^{6}$D$^{\rm o}$ & 6.15 & 3/11 \\
  29 & a$^{4}$F-z$^{6}$D$^{\rm o}$ & 6.15 & 1/11 \\
\hline
  20 & b$^{4}$D-z$^{4}$F$^{\rm o}$ & 6.41 & 1/9  \\
  30 & a$^{4}$F-z$^{4}$F$^{\rm o}$ & 6.41 & 3/10 \\
  50 & b$^{4}$G-z$^{4}$F$^{\rm o}$ & 6.41 & 4/7  \\
\hline
  26 & b$^{4}$P-z$^{4}$D$^{\rm o}$ & 6.76 & 1/7  \\
  31 & a$^{4}$F-z$^{4}$D$^{\rm o}$ & 6.76 & 4/9  \\
  44 & b$^{4}$F-z$^{4}$D$^{\rm o}$ & 6.76 & 6/9  \\
\hline
 117 & b$^{2}$G-z$^{4}$G$^{\rm o}$ & 8.07 & 2/3  \\
\hline
\end{tabular}
\end{center}
\end{table}

\section{Analysis}

\subsection{Line fitting}

 All spectra have been deredshifted using z=0.149. The lines were fit using a 
code originally written by E. Zuiderwijk and described in V\'eron et al. 
(\cite{veron80}). 

 We started by fitting the Keck spectrum, which has the largest spectral range of all 
available spectra, although it does not have the best spectral resolution nor the best
S/N ratio. We used only the spectral range 3450-6960 \AA, the red end of the spectrum 
being noisy. We masked the range 6610-6680 \AA\, which is affected by the atmospheric 
A band. The other three spectra were then fit with the same set of emission lines.

 Initially, Gaussian profiles were adopted. To get a good fit three emission 
line systems with different line velocity and width were required : a very weak 
narrow line (NLR), a broad line 
(BLR) and a very broad line (VBLR) systems. In each system all the lines were
forced to have a Gaussian profile with the same velocity and width. The resulting 
low value of the BLR FWHM, $\sim$1\,780 km s$^{-1}$, qualifies this object to be classified as a NLS1.

 However, one difficulty with this fit is that a few of the strongest emission 
lines from Fe\,II 
multiplets 27, 42 and 49 (especially m. 42 $\lambda$5018 and $\lambda$5169) appear 
to have a narrow core (1\,080 km s$^{-1}$ FWHM) at the velocity of the broad line 
system (fig. \ref{SP_IRAS}). \\

As V\'eron-Cetty et al. 
(\cite{veron01}) have shown that the broad emission lines of NLS1s are better fit
with Lorentzians than with Gaussian profiles, we repeated the fitting process using Gaussian 
profiles for the VBLR and Lorentzians for the BLR. This fit was not satisfactory as 
the continuum was unacceptably low, especially in the region $\lambda\lambda$
5600-5800 \AA\ where few emission lines are expected. This is due to the fact that 
the BLR spectrum contains a large number of lines. Lorentzian profiles have extended 
wings; the addition of the wings of all these lines produces a pseudo continuum which 
pushes down the true continuum. It is possible that the true profiles could be
Lorentzians with truncated wings, as very high velocity extensions of the emitting clouds are unlikely (There 
is  no reason to believe that the line profile is strictly Lorentzian or Gaussian).
To test this hypothesis, we fit the BLR Balmer lines and the strongest metallic lines 
with Lorentzians and the other with Gaussians, with all lines of this system having the 
same central velocity. The resulting fit is very similar to the one obtained with   
Gaussians only, except for two differences: a) the NLR has vanished which suggests that it 
is not real and b) the narrow core in the strongest metallic lines has disappeared.

 Using only Gaussian profiles, we obtained the following 
results:

1/ The BLR has a FWHM $\sim$1\,780 km s$^{-1}$. In addition a second component was 
needed to fit the H$\alpha$ line. This additional component is blueshifted by $\sim$1\,300 km s$^{-1}$ with 
respect to the 
main component and has a FWHM of $\sim$1\,000 km s$^{-1}$. Its flux is equal to 
$\sim$7\% of the flux of the stronger component. Boroson \& Meyers (\cite{boroson92}) 
have shown that the BALQSOs have an H$\alpha$ line with a large blue asymmetry
similar to the one observed here. 

\begin{figure*}[ht]

\resizebox{17.5cm}{!}{\includegraphics{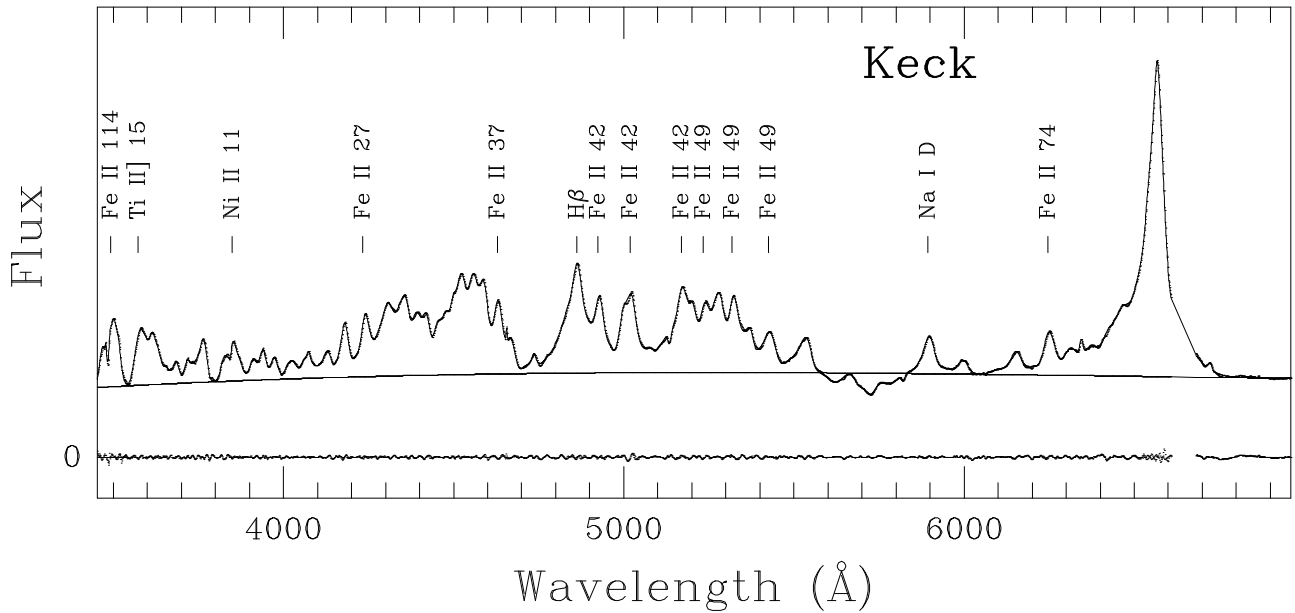}}

\caption{\label{SP_OGLE} The deredshifted Keck spectrum of IRAS\,07598+6508 with the 
fit, the continuum and the residuals. The strongest emission lines have been identified.} 

\end{figure*}

 The FWHM of the H$\alpha$ line in the BLR has been estimated to be 2\,550 km 
s$^{-1}$ (Boroson \& Meyers \cite{boroson92}) and that of H$\beta$ 3\,200 
km s$^{-1}$ (Lipari et al. \cite{lipari93}), 3\,150 km s$^{-1}$ (Zheng et al. 
\cite{zheng02}) or even 4\,850 km s$^{-1}$ (Marziani et al. \cite{marziani03}). The 
significantly smaller width found here may be attributed to the identification of a very broad component. 
The quoted values must refer 
to the complex profile made of the two broad components. 
 
2/ The VBLR ($\sim$7\,500 km s$^{-1}$ FWHM) is blueshifted by $\sim$ 760 km s$^{-1}$ 
with respect to the BLR. Its flux is about three times larger than that of the BLR. 
 Note that the
parameters of this component are rather poorly determined due to the presence of the 
atmospheric A band in the H$\alpha$ red wing.
Such a VBLR seems to be common in QSOs, blueshifted by $\ge$ 1\,000 km s$^{-1}$, 
with a width $\ge$ 7\,000 km s$^{-1}$ (Brotherton et al. \cite{brotherton94}). \\

 Fig. \ref{ERR1_IRAS} shows plots of the logarithm of the relative peak intensity 
of all detected lines in the Bok 2.3-m spectrum {\it vs} the corresponding values in 
the KPNO 4-m and Keck 10-m spectra respectively. It appears that the accuracy is of
the order of 20 \% for the strongest lines and about a factor of 2 for the weakest. \\

\begin{figure}[ht]

\resizebox{8.8cm}{!}{\includegraphics{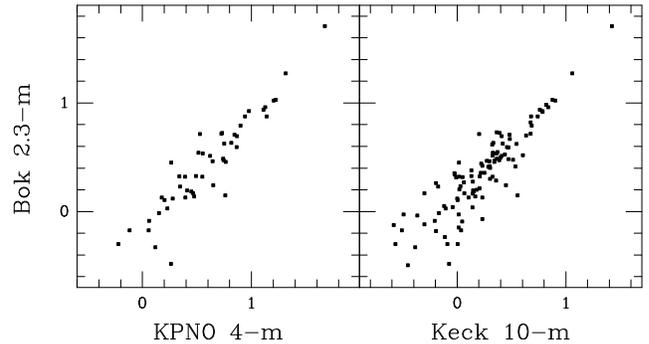}}

\caption{\label{ERR1_IRAS} Plot of the logarithm of the relative peak intensity of the 
detected lines in the Bok 2.3-m spectrum {\it vs} the corresponding values in the KPNO 
4-m and Keck 10-m spectra respectively.} 

\end{figure}

 The best fit of the Keck spectrum is shown in figure \ref{SP_OGLE}. \\

 The P$\alpha$ line shows a blueshifted broad component (FWHM$\sim$3\,900 km s$^{-1}$)
together with a narrow core (FWHM$\sim$530 km s$^{-1}$) (Taniguchi et al. \cite{taniguchi94}).
The broad component could be produced by both the BLR and the VBLR.

 Hines \& Wills (\cite{hines95}) found that the peaks of the high-ionization UV lines 
are blueshifted with respect to the H$\alpha$ and Na\,I\,D $\lambda$5892 emission peaks 
by 3\,000 km\,s$^{-1}$. It is difficult to associate these lines with any of the three 
systems listed above as none of them have such a large blueshift with respect to the broad 
line system.

\subsection{The emission spectrum}

\subsubsection{No narrow line component (NLR) ?}

 As shown above, the weak narrow line component may not be real, in agreement
with Boroson \& Meyers (\cite{boroson92}). However weak [O\,III] lines have been 
observed with $\lambda$5007/H$\beta\sim$0.02 by Lipari (\cite{lipari94}). 

\subsubsection{The broad line component (BLR)}

 Due to the large number of emission lines and their substantial width, most 
of them are heavily blended. Consequently, it is impossible to identify them all unambiguously. 
Therefore, our process relies on some preconceived ideas. 
 As it is well known that Fe\,II is an important contributor, we have included in our 
model all the strongest Fe\,II lines appearing in high density, high optical thickness 
models. Such line lists can be obtained by running the code CLOUDY with the large Fe$^{+}$
atom (Ferland \cite{ferland02}).

 As the ionisation potentials of Fe, Ti and Cr are very similar (6.8-7.8 eV for 
the neutral atoms and 13.6-18.4 eV for the once ionized ions), lines of all these 
elements are expected to be present. It is indeed the case in the spectrum of the 
very peculiar star XX\,Ophiuci (Merrill \cite{merrill51}; \cite{merrill61}; Cool et 
al. \cite{cool05}) nicknamed the "iron star" by Merrill (\cite{merrill24}). Therefore, 
we include in our analysis all lines of these elements observed in the spectrum of XX\,Oph.

 It turns out that most of the multiplets of Fe\,II, Ti\,II and Cr\,II observed
in XX\,Oph are indeed present in the spectrum of IRAS\,07598+6508 (they are listed in 
tables \ref{tfe2}, \ref{tti2} and \ref{tcr2}, for Fe\,II, Ti\,II and Cr\,II respectively).
Two emission features could not be identified (at $\lambda$5051 and $\lambda$6339).

 In addition to these metallic lines, the Balmer lines (H$\alpha$ to H$\delta$), 
Ca\,II $\lambda\lambda$3933,3968 and Na\,I\,D $\lambda\lambda$5890,5896 are present. 
Na\,I\,D has previously been observed in emission with EW=9.7 \AA\ (Boroson \& Meyers 
\cite{boroson92}) and Na\,I\,D/H$\alpha$$\sim$0.05 (Lipari \cite{lipari94}). We found 
EW(Na\,I\,D)=9.6-10.3 \AA\ and Na\,I\,D/H$\alpha$$\sim$0.16-0.20, but this ratio uses 
the flux of the broad H$\alpha$ component rather than the total H$\alpha$ flux,
which probably explains why we found a much larger value for the Na\,I\,D/H$\alpha$ 
ratio.

 Lawrence et al. (\cite{lawrence97}) found Fe\,II $\lambda$4570/H$\beta$$\sim$2.6, 
Lipari et al. (\cite{lipari93}) $\sim$2.6, Lipari (\cite{lipari94}) $\sim$2.3, 
Zheng et al. (\cite{zheng02}) 2.75 and Marziani et al. (\cite{marziani03}) 1.21. 
These values are in good agreement, except for the last one which is about twice smaller; 
however, our 
value is much larger, on the order of 8. This is due to the fact that the H$\beta$ 
flux in the BLR is only a quarter of the total H$\beta$ flux as we have seen above.
Indeed we compare the Fe\,II flux to that of the H$\beta$ flux coming from the same 
region, namely the BLR.

 A Balmer line ratio H$\alpha$/H$\beta$$\sim$5-6.2 was measured, suggesting the 
presence of reddening in the emission line region (Lipari \cite{lipari94}; Hines \& 
Wills \cite{hines95}), although this could be explained by radiative transfer effects 
in high optical thickness media (Collin-Souffrin et al. \cite{collin82}). For the BLR 
we found that this ratio is $\sim$6.3. The spectral energy distribution is not reddened 
and matches that of typical QSOs (Hines \& Wills \cite{hines95}). \\

 The list of all lines detected in the BLR in each of the four available spectra is 
given in Appendix A. All these lines have been fit with a Gaussian profile having
the same velocity width (1\,780 km\,s$^{-1}$ FWHM).

\subsubsection{The very broad line component (VBLR)}

 From this region, we detect only the Balmer lines from H$\alpha$ to H$\delta$.

\subsection{The Na\,I D absorption}

 A narrow Na\,I\,D doublet has been seen in absorption at an outflow velocity of 3\,800 
km s$^{-1}$ and EW=0.95 \AA\ (Boroson \& Meyers \cite{boroson92}). It is clearly seen 
at that velocity and strength in the KPNO 4-m spectrum. Rupke et al. (\cite{rupke05}) 
have observed this doublet at V=--3\,939 $\pm$10 km s$^{-1}$, with FWHM=130 km s$^{-1}$.

 In addition, a broad absorption trough is seen from about $\lambda$5550 \AA\ to 
about $\lambda$5800 \AA\ in the object rest frame, corresponding to outflow velocities 
from the narrow system up to about 16\,000 km s$^{-1}$ (Boroson \& Meyers 
\cite{boroson92}). We have modelled this broad absorption feature with four Gaussians 
blueshifted  by 14\,630, 11\,300, 9\,400 and 6\,340 km s$^{-1}$,  having FWHMs of 2\,900, 1\,250, 2\,360 and 3\,100 km 
s$^{-1}$ and EWs of 5.9, 2.2, 10.0 and 7.9 \AA\ respectively.
 The general pattern of the absorption is quite similar to what is seen in Mg\,II 
(Lipari \cite{lipari94}). Strong high-ionization BALs have also been observed in
this object (Lipari \cite{lipari94}; Turnshek et al. \cite{turnshek97}). 

 The profile of the broad absorption feature appears quite similar in the four 
spectra we have analysed, taken in 1991 and 1995. However, Rupke et al. have obtained 
on April 13, 2004 a high S/N ratio spectrum with the KPNO 4-m telescope (5400 s 
exposure), with a resolution of 85 km s$^{-1}$. The profile of the absorption feature 
is quite different in that spectrum; the bluest (--14\,630 km s$^{-1}$) and the reddest 
(--6\,340 km s$^{-1}$) components have completely disappeared. 

 A Galactic Na\,I\,D doublet is present with an EW of 0.65\,\AA.
  
\section{Discussion}

\subsection{The theoretical Fe\,II emission spectrum}

 Baldwin et al. (\cite{baldwin04}) showed that photoionized BELR clouds cannot produce
both the observed shape and equivalent width of the 2200-2800 \AA\ Fe\,II UV bump in 
active galactic nuclei, unless there is a considerable velocity structure 
corresponding to a  microturbulent velocity parameter v$_{\it turb}>$ 100 km s$^{-1}$ 
for the emitting cloud. An alternative solution is that the Fe\,II emission is the result 
of collisional excitation in a warm, dense gas. However, they show that gas with 
temperature 6\,000\,K $<$ T $<$ 40\,000\,K, density n$_{H}\sim$ 10$^{12}$-10$^{16}$ cm$^{-3}$
and column density N$_{H}$$\sim$ 10$^{25}$ cm$^{-2}$ will emit primarily Fe\,II UV
lines. Since this gas does not emit strongly in lines of other elements, it would have 
to constitute another component in an already complicated picture of the BELR and
consequently these authors prefer the model involving turbulence. \\

 We will show here that photoionization models are actually not able to explain 
the observed emission lines of IRAS\,07598+6508, while some hope arises from purely 
collisional models.

 Using the code CLOUDY with its large Fe$^{+}$ atom (Ferland \cite{ferland02}) we 
computed a number of models to match the observed Fe\,II spectrum and the main 
BLR features. Unfortunately CLOUDY does not provide information on Ti\,II, Ni\,II 
or Cr\,II optical lines. The intensity of the Fe\,II predicted 
lines can be obtained either separately or summed over wavelength bands directly 
comparable to the observations. 

 The discussion is summarized in Table \ref{model}, which gives in the first column 
the main features observed in the BLR, in column 2 the wavelength of the 
lines or the wavelength range for the Fe\,II bands, in columns 3, 4, 5 and 6 the line 
intensity ratios referred to H$\beta$ measured in the four available spectra, and 
in the last five columns the results from different models. For each of these models 
several attempts have been made to reproduce the observations, but we only give 
here the best fits.

 Our first attempt was to compute a standard photoionization model, assuming that the 
size of the emission region follows the relationship between the size of the BLR and 
the luminosity of the central source of radiation deduced by Kaspi et al. 
(\cite{kaspi00}) from reverberation mapping of a sample of AGN. Assuming an optical 
luminosity $\sim$10$^{45}$ erg\,s$^{-1}$, we infer a distance of the BLR from the 
central source of radiation of $\sim$4$\times$10$^{17}$ cm. The results of a model with 
a density n=$10^{12}$ cm$^{-3}$ and a column density $N_H$=2$\times$$10^{23}$ cm$^{-2}$,
values commonly adopted for the BLR, are given in column 7 of Table \ref{model} 
(model 1). It clearly shows some problems: the excess strength of the predicted He\,I, 
Ly\,$\alpha$, Mg\,II and Balmer continuum, as well as the weakness of the Fe\,II 
features. The former lines are formed in the H\,II region of the irradiated cloud, but 
no such strong features are observed in the UV spectrum of IRAS\,07598+6508 (cf. 
Lanzetta et al. \cite{lanzetta93} and Lipari \cite{lipari94}) or of any AGN, and no 
He\,I $\lambda$5876 is identified in any of the four optical spectra.

\begin{table*}[!t]
\caption{\label{model}Observed and computed line ratios in IRAS\,07598+6508. A "--" 
indicates that the line is outside the spectral range or in a region of the spectrum 
masked because of the presence of poorly corrected atmospheric features.}
\bigskip
\begin{tabular}{|c|c|r|r|r|r|r|r|r|r|r|r|}
\hline
\noalign{\smallskip}
lines &$\lambda$& Keck&KPNO&KPNO&Bok & model 1& model 2& model 3 & model 4 & model 5\\
      &         &     &4m &2.1m&2.3m&phot. & phot.  & phot. & coll.& coll.\\
      &         &     &&&&``Kaspi''    & +heating  & +heating &  &v=100    \\
      &         &  &&&&R=4$\;10^{17}$&R=3$\;10^{19}$&R=5$\;10^{18}$&f=3\%&f=5\%\\
      &     &&&&&n=$10^{12}$&n=$10^{12}$&n=$10^{14}$&n=$10^{15}$&n=$10^{15}$\\
      &        &&&&&&H=6$\;10^{45}$&H=2$\;10^{46}$&H=6$\;10^{44}$&H=$6\;10^{44}$\\
      &&&&&&N$_H$=2$\;10^{23}$&N$_H$=$10^{23}$&N$_H$=5$\;10^{23}$&N$_H$=$10^{24}$&N$_H$=$10^{25}$\\
\noalign{\smallskip}
\hline
\noalign{\smallskip}

H$\alpha$&      & 5.86 &6.71&4.28&8.18 & 3.25& 7.83 & 4.5  & 0.34 &0.12 \\
H$\beta $&      & 1.00 &1.00&1.00&1.00 & 1.00& 1.00 & 1.00 & 0.01 &0.02 \\
H$\gamma$&      & 0.29 &--  &0.28&0.21 &0.52 & 0.53 & 0.37 & 0.00 &0.00 \\
H$\delta$&      & 0.15 &--  &0.07&0.08 &0.40 & 0.36 & 0.18 & 0.00 &0.00 \\

         &      &     & &  &   & &      &      &  &   \\
Ca\,IIK  & 3934 & 0.22&--  &0.28&--  &0.63 & 2.76 & 16.5 & 1.34 &0.88 \\
Ca\,IIH  & 3969 & 0.19&--  &0.29&0.14&0.46 & 2.05 & 14.7 & 1.28 &0.78 \\
He\,I    & 5876 & 0.00&0.00&0.00&0.00&2.62 & 0.21 & 0.19 & 0.00 &0.00 \\
Na\,I    & 5892 & 0.87&1.00&0.75&1.00&0.18 & 6.60 & 10.9 & 0.65 &0.50 \\

         &         &   &   &&&      &      &      &   &  \\
Fe\,II   &         &   &   &&&      &      &      &   &  \\
3590    &3400-3780&$>$1.57&--       &1.35    &--     & 0.60 & 4.19 & 5.91 & 6.04 &6.32 \\
3910     &3780-4040& 1.23 &--       &1.54    &--     & 0.74 & 6.71 & 11.5 & 5.62 &5.84 \\
4060     &4040-4080& 0.19 &--       &0.10    &0.11   & 0.01 & 0.02 & 0.03 & 0.13 &0.16 \\
4255     &4080-4430& 5.64 &--       &6.57    &5.24   & 0.60 & 28.3 & 6.51 & 7.40 &7.52 \\
4570     &4430-4685& 7.91 &--       &8.59    &7.89   & 0.84 & 10.0 & 9.49 & 8.00 &8.00 \\
4743     &4685-4800& 0.21 &--       &0.91    &0.29   & 0.03 & 2.22 & 0.51 & 0.49 &0.48 \\
4855     &4800-4910& 1.27 &--       &2.48    &1.44   & 0.07 & 6.68 & 1.86 & 1.09 &1.12 \\
4975     &4910-5040& 4.23 &$>$3.12  &7.49    &4.04   & 0.52 & 9.76 & 4.90 & 1.42 &1.44 \\
5070     &5040-5100& 0.00 &0.00     &0.00    &0.00   & 0.02 & 0.55 & 0.34 & 0.27 &0.28 \\
5143     &5100-5185& 3.26 &4.60     &3.31    &3.16   & 0.28 & 11.0 & 3.68 & 2.18 &2.32 \\
5318     &5185-5450& 8.20 &11.69    &8.82    &8.80   & 0.65 & 19.1 & 7.11 & 5.19 &5.36 \\
5540     &5450-5630& 0.80 &1.07     &0.65    &0.88   & 0.09 & 3.01 & 1.24 & 1.03 &1.04 \\
5865     &5630-6100& 0.77 &0.62     &0.58    &0.09   & 0.10 & 1.65 & 1.92 & 1.39 &1.44 \\
6265     &6100-6430& 3.93 &4.46     &4.79    &3.50   & 0.12 & 0.68 & 1.86 & 2.25 &2.32 \\
6565     &6430-6700& 1.15 &1.98     &4.79    &1.79   & 0.25 & 2.55 & 4.55 & 1.77 &1.84 \\
6910     &6700-7120& 0.00 &--       &--      &--     & 0.02 & 0.49 & 0.03 & 0.12 &0.12 \\
7445     &7120-7770&   -- &--       &--      &--     & 0.15 & 6.47 & 1.73 & 2.69 &2.72 \\

          &         &     &&&&      &      &      &   &   \\
          &         &     &&&&      &      &      &   &  \\
L$\alpha$ &1216               &  --  &--&--&--& 34.4 & 37.0  &  1.2 &  0.0\,\, &1.12  \\
Fe\,II2355&2280-2430          &  --  &--&--&--&  4.1 & 21.0  & 15.4 & 11.6\,\, &6.48  \\
Mg\,II    & 2800              &  --  &--&--&--& 11.4 &198.0  & 13.3 &  1.44    &1.20  \\
Ba\,C     &$<$3646\,\,\,\,\,  &  --  &--&--&--& 55.5 &  2.7  &  1.0 &  0.04    &0.03  \\
CaII\,T   & 8500              &  --  &--&--&--&  1.4 & 11.8  & 48.0 &  2.96    &3.68  \\

\noalign{\smallskip}
\hline
\end{tabular}
\end{table*}

 A way to weaken the influence of the H\,II region is to locate the BLR farther away 
from the central source. A distance as large as 3$\times$$10^{19}$ cm is needed
to lower the He\,I $\lambda$5876/H$\beta$ ratio down to 0.2. Such a model has the
disadvantage of not producing enough Fe\,II emission in addition of being difficult 
to explain in term of the location of the BLR.

 Putting aside this distance problem for a while, it is possible to strengthen the 
emission of the excited H\,I$^*$ region if an additional heating mechanism, such as 
a mechanical one, is at work. The results of such a model are given in column 8 
(model 2), where the additional heating is assumed to be of the order of the bolometric 
luminosity (H$_{extra}$ = $6\times10^{45}$ erg s$^{-1}$). Although it reproduces 
the overall Fe\,II intensity, the line ratios are not in agreement with the 
observed ones; this is, in particular, the case for Mg\,II which, in this model, is 
inconsistent with the observations.

 An increase of the density of the medium up to $10^{14}$ cm$^{-3}$, as  proposed in 
column 9 (model 3), produces a substantial decrease of Mg\,II but also worsens the match to the Ca\,II and 
Na\,I\,D line intensities. Note however that high densities not only produce Fe\,II 
line ratios in good agreement with the observations, but that they also require the BLR to be located 
closer to the central source, necessitating in addition a very strong contribution from mechanical heating. 

  Faced with the undeniable difficulties of the models where the primary source 
of excitation is radiative -even if the addition of another source of heating 
improves the result- we turned to purely collisional models, $\it i.e.$ models where 
the emission region is shielded from the central source of radiation and where the 
source of excitation is only due to mechanical heating. Mechanical heating was first 
proposed by Collin-Souffrin (\cite{coll86}) to solve the energy puzzle in the BLR. 
This heating can be produced in the accretion disc around the massive black hole by 
the interaction between accreting matter and magnetic field (Kwan et al. \cite{kwan95}, 
Hirotani et al. \cite{hirotani92}). It can also be produced in the atmosphere of the 
accretion disc by viscous energy release (Blaes et al. \cite{blaes01}). Similarly, 
Lipari et al. (\cite{lipari05}) suggested that Fe\,II emission could originate in 
warm regions obscured from the direct ionizing UV photons, the obscuring material 
being in the form of expanding shells. In particular, they classify IRAS\,07598+6508 
as a BAL IR AGN associated with strong early starburst activity. The giant explosive 
events occuring from the evolution of very massive star would produce shock-heated 
material.

  In view of the huge Fe\,II emission observed in IRAS\,07598+6508 and of the weakness 
of all other lines, except Ti\,II and Cr\,II which actually have similar 
ionization potential, our aim is to find a region which would emit Fe\,II lines and 
no others (or, at least, which would emit very weak H\,I, He\,I, 
Na\,I, Ca\,II and Mg\,II). It is known since the works of House (\cite{house64}) and 
Jordan (\cite{jordan69}) that, in a collisional medium where the radiation field is 
negligible, each ion is emitted at a specific temperature and therefore in a specific
region. The relative importance of these regions strongly depends on the distribution 
of temperature, and therefore, on the column density of the cloud (as the imposed 
parameter here is the total heating).

 The results obtained with such models are displayed in the last two columns of 
Table \ref{model}, where H is the assumed mechanical heating in erg s$^{-1}$. As 
very weak emission from the Balmer lines is expected, the line ratios have to be 
considered relative to the Fe\,II intensity rather than to H$\beta$ as previously. 
We used the Fe\,II $\lambda$4570 blend as a reference, and its predicted intensity 
is scaled to the observed one averaged over the four available spectra 
($\sim$1.7$\times$10$^{-13}$ erg s$^{-1}$ cm$^{-2}$). From this scaling, we determine 
the emission surface, {\it i.e.} the covering factor f of the emission region 
(given in Table \ref{model}, assuming a radius of the BLR of a few times 
10$^{17}$ cm). The covering factor is determined by comparison between the
luminosity in the Fe\,II $\lambda$4570 blend and that computed by CLOUDY, assuming a 
radius for the BLR in agreement with the Kaspi relation(Kaspi et al. \cite{kaspi00}). 
The computed luminosity of Fe\,II is also a function of the mechanical heating, 
which is arbitrarily chosen so that the input energy is not larger than the 
bolometric luminosity. This input energy is mainly determined by the line ratios. 
The source of the mechanical heating has yet to be determined.

 A relatively good fit was obtained for a high density (n=$10^{15}$ cm$^{-3}$), 
high column density (N$_H$=$10^{24}$ cm$^{-2}$) model (model 4). The density 
increase has the advantage of 
decreasing the relative intensity of Na\,I, Ca\,II, Mg\,II and Fe\,II UV with 
respect to Fe\,II $\lambda$4570. A further increase of the column density would 
decrease even more Na\,I, Ca\,II, Mg\,II but it would increase Fe\,II UV. The 
Fe\,II UV feature at 2355 \AA\ is one of the features that dominate the computed 
spectra over much of the parameter space in the models discussed by Baldwin et 
al. (\cite{baldwin04}). In that paper the authors call for microturbulence to 
improve the situation. Even in the context of a non-photoionized model, 
microturbulence can transfer, through line and continuum fluorescence, a 
fraction of the near UV emission into the optical range, improving the agreement 
with observed line ratios. A small improvement is actually obtained assuming a 
microturbulence $v_{turb}$=100 km s$^{-1}$. In particular, the pumping of the 
Fe\,II UV lines allows increasing the column density up to N$_H$=$10^{25}$ 
cm$^{-2}$, inducing a small decrease of Mg\,II, Ca\,II\,H and Ca\,II\,K (model 5).

 The Fe\,II line ratios show some disagreements: the predicted 
Fe\,II $\lambda$3590 and $\lambda$3910 are much too strong, and Fe\,II $\lambda$4975 
is much too weak.

 The Fe\,II $\lambda$4975 band mainly includes two lines of multiplet 42. It 
appears that the atomic data base of CLOUDY has no collisional strength listed for 
these two lines, and so, consequently, a very small one is used in the computations. Not 
only in IRAS\,07598+6508 are these lines observed with a large intensity, but also in 
most Fe\,II emitters, suggesting that their collisional strength is probably not
negligible. The underestimation of the collisional strengths may be the explaination 
for the discrepancy.

\begin{figure}[h]

\resizebox{8.8cm}{!}{\includegraphics{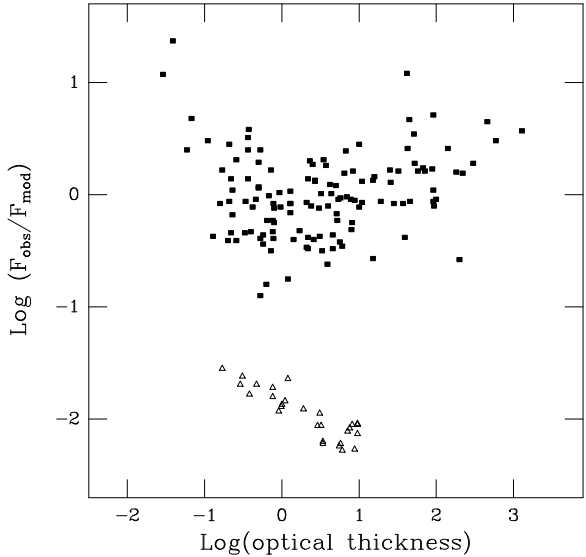}}

\caption{\label{CORREP} This figure shows the ratios between the observed and
predicted line intensities $\it vs$ the optical thicknesses (filled sqares).
Also shown are the unobserved lines predicted to have a strong intensity in the
model (open triangles).}

\end{figure}

 The high intensity of the two features, Fe\,II $\lambda$3590 and Fe\,II 
$\lambda$3910, is due to the intercombination multiplets 3, 4, 10, 14, 22 and 23, 
which are all predicted to be strong in the models, but are conspicuously absent 
in the observed spectra. Whatever the physical parameters assumed for the BLR, these 
two features are at least as strong as half the value for Fe\,II $\lambda$4570. Fig. 
\ref{CORREP} shows the ratios between the observed and the predicted Fe\,II line 
intensities $\it vs$ the optical thicknesses. Fe\,II lines which are strong in the models 
but could not be detected are shown as upper limits (open triangles). To  produce this figure 
we use the potentiality given by CLOUDY for obtaining the intensity of each Fe\,II line
separately. We have no obvious answer for this problem. We note that it concerns only 
lines with a moderate transition probability and therefore a moderate optical thickness, 
but not all lines with such characteristics, and, in particular, not the lines in the red. 
Indeed the puzzling lines decay from levels around 5 eV down to very low levels (below 
2.5 eV).

 Two possible explanations can be invoked: again the lack of collisional strengths between some 
of these low levels, and the escape probability approximation used in CLOUDY. Avrett 
\& Loeser (\cite{avrett87}) have shown that using the escape probability approximation 
in multilevel problems instead of solving the transfer equations can induce large
errors for all line transitions and, in particular, for the weak ones (see also 
Collin-Souffrin \& Dumont \cite{collin86}).

\subsection{The Fe\,I emission spectrum}

 Bergeron \& Kunth (\cite{bergeron80}) suggested the presence of Fe\,I emission 
in the spectrum of PHL\,1092. Three individual features at $\lambda$3575, 3763 
and 3851 were the clearest indications for the presence of this ion. The best 
identified multiplets were 4, 5, 6, 20, 21, 23, 24 and 45. All the lines that 
are strong in the laboratory were present. All the identified multiplets arise 
from upper levels at less than 4.8 eV.

 Kwan et al. (\cite{kwan95}) have identified Fe\,I and Ti\,II lines, in addition 
to Fe\,II, in the spectrum IRAS\,07598+6508. They have noted three regions in 
the spectral range 3050-4570 \AA\ whose features cannot be accounted for by Fe\,II 
lines. The first region is from 3315 to 3450 \AA\ (outside our observed range),
the second from 3520 to 4090 \AA, the third from 4400 to 4480 \AA. Kwan et al. 
suggested that the emission in these three spectral intervals is mainly due to 
Fe\,I lines. \\

 Sigut et al. (\cite{sigut04}) have predicted the Fe\,I emission in the BLR in 
various models. The strongest transitions are decays from low-lying odd parity 
levels between $\sim$3.5 to 4.5 eV to the three lowest even parity states, the 
a$^{5}$D$^{\rm e}$ ground state, a$^{5}$F$^{\rm e}$ and a$^{3}$F$^{\rm e}$. The strongest 
predicted multiplet is 23, z$^{5}$G$^{\rm o}$-a$^{5}$F$^{\rm e}$, giving lines near 
$\lambda$3600 \AA. The models computed have N$_{e}$=10$^{9.6}$ and 10$^{10.6}$ 
cm$^{-3}$, and a column density N$_{H}$=10$^{23}$ cm$^{-2}$. The strongest 
predicted Fe\,I lines occur for the highest value of the electron density.
However the predicted fraction of Fe\,I in all computed models is very small 
($<$10$^{-3}$): at large optical depth, Fe$^{+}$ is always the dominant iron 
species. 

 Very few Fe\,I lines are computed by CLOUDY. Their strength is always much lower 
than that of the Fe\,II lines ($\sim$10$^{-2}$), except for the hybrid model 
(radiative plus mechanical heating) of high density (cf. model 3, column 9 of 
Table \ref{model}), where the temperature is less than 6000\,K; but we have seen 
that this model otherwise produces inconsistent results.

 The flux of the Fe\,I lines identified by Bergeron \& Kunth (\cite{bergeron80}) 
and Kwan et al. (\cite{kwan95}) in the spectrum of PHL\,1092 and IRAS\,07598+6508 
relative to the nearby Fe\,II flux between 3600 and 3800 \AA\ is much 
larger than predicted by these models. As we have seen above, we were able to get 
a good fit by using instead of Fe\,I lines, Ti\,II and Cr\,II lines from multiplets 
observed in the spectrum of XX\,Oph.

\subsection{Comparison of the emission line regions in I\,Zw\,1 and IRAS\,07598+6508}

\subsubsection{The broad line region} 

 In a recent paper (V\'eron-Cetty et al. \cite{veron04}), we made a detailed analysis 
of the emission spectrum of the NLS1 I\,Zw\,1. We shall stress here the similarities and 
differences between the spectra of I\,Zw\,1 and IRAS\,07598+6508.

 One of the main differences between these two objects is the much larger strength 
of the metallic lines with respect to the Balmer lines in IRAS\,07598+6508. The 
parameter R$_{4570}$ ($\it i.e.$ the ratio of the line flux in the BLR in the range 
4430-4685\,\AA\ to the H$\beta$ flux from the same region) is $\sim$8 in 
IRAS\,07598+6508, as compared to 1.7 in I\,Zw\,1. Moreover, the relative Fe\,II 
line intensities are significantly different in the two objects. For instance, the
intercombination multiplets 16, 25, 26, 29, 35 and 36 are absent or weak in I\,Zw\,1, 
while they are relatively strong in IRAS\,07598+6508. The Na\,I\,D lines are five 
times stronger relative to H$\beta$ in IRAS\,07598+6508 than in I\,Zw\,1, $\it 
i.e.$ their intensities relative to the Fe\,II lines are the same in the two objects. 
We had not detected the Ca\,II H and K lines in I\,Zw\,1. He\,I and Si\,II lines 
were observed in I\,Zw\,1; they are absent from IRAS\,07598+6508. 

 In I\,Zw\,1 we tentatively identified 27 lines with high-excitation Fe\,II lines. 
We had then not yet recognized the importance of Ti\,II and Cr\,II. Today we would 
probably identify many of them with lines from these two ions and so the two metallic 
spectra are qualitatively but not quantitatively similar.

 Although the differences in Fe\,II line intensities are substantial, the excitation 
mechanism is probably the same. Indeed we had shown that a standard photoionization 
model was not able to account for the Fe\,II strength observed in the BLR of I\,Zw\,1
(V\'eron-Cetty et al. \cite{veron04}). Difficulties similar to the ones we encounter 
here occur, such as huge predicted Ly\,$\alpha$, Mg\,II, Balmer continuum as well as 
He\,I lines in excess of the observations. If again, we 
adopt mechanical heating alone for the excitation, very good agreement with the 
observations is obtained (except for Fe\,II $\lambda$3910) with a density n=$10^{14}$ 
cm$^{-3}$, a column density N$_H$=$10^{24}$ cm$^{-2}$ and a heating H=2$\times$10$^{44}$ 
erg s$^{-1}$. To account for the observed luminosities a covering factor f=5\,\% is 
necessary. These characteristics of the emission region are very close to that obtained 
for IRAS\,07598+6508, although the higher intensity of Fe\,II relative to H$\beta$ in 
the latter object, together with the increase of the relative intensities of some
intercombination mutiplets, imply a somewhat higher density and column density
(n=$10^{15}$ cm$^{-3}$ and N$_H$=$10^{25}$ cm$^{-2}$).

\subsubsection{The narrow line regions} 

 In the spectrum of I\,Zw\,1 we had found a very rich low excitation NLR together 
with two high excitation NLR. Nothing similar was found in IRAS\,07598+6508.

\section{Conclusion}

 We have shown that the emission line spectrum of the NLS1 Seyfert galaxy IRAS\,07598+6508
is dominated by a BLR region emitting mainly, apart from the Balmer lines, Fe\,II, Ti\,II
and Cr\,II lines. The best model accounting for this BLR is a purely collisional model with 
a high density (n=10$^{15}$ cm$^{-3}$), a high column density (N$_{H}$=10$^{24}$ cm$^{-2}$), 
a microturbulence of 100 km s$^{-1}$, a mechanical heating of the order of one tenth of the
bolometric luminosity ($\sim$10$^{45}$ erg s$^{-1}$ and a low covering factor (f=5\,\%). This 
BLR is qualitatively similar to the one observed in I\,Zw\,1, but we have not found traces 
in IRAS\,07598+6508 of the narrow line regions found in the latter object.

\section{Acknowledgments}

 We gratefully thank 
Gary Schmidt, who kindly put at our disposal his spectrum of IRAS\,07598+6508, and 
S. Collin for helpful discussions.

{}
 \newpage
 \appendix

\begin{table*}[!ht]
\caption{\label{synops}Appendix A. 
Observed line ratios (relative to H$\beta$) in the IRAS\,07598+6508 BLR. The H$\beta$ 
flux is $\sim$2.3$\times$10$^{-14}$ erg s$^{-1}$ cm$^{-2}$. An "n" indicates that 
the line has not been detected, an "h" that it is outside the spectral range or in a 
region of the spectrum masked because of the presence of poorly corrected atmospheric 
features.}
\begin{center}
\begin{tabular}{|l|c|c|c|c|c|}
\hline
 Line & $\lambda (\AA)$ & KPNO 4-m & 2.3-m Bok & Keck & KPNO 2.1-m \\
\hline
 H$\beta$   & 4861.30 &     1.00 &     1.00 &     1.00 &     1.00  \\
 H$\alpha$  & 6562.80 &     6.71 &     8.18 &     5.87 &     4.28  \\
 H$\gamma$  & 4340.40 &      h   &     0.21 &     0.29 &     0.28  \\
 H$\delta$  & 4101.74 &      h   &     0.08 &     0.15 &     0.07  \\
 Ti II 6    & 3444.31 &      h   &      h   &      h   &     0.32  \\
 Ti II 6    & 3461.50 &      h   &      h   &     0.15 &     0.12  \\
 Ti II 99   & 3465.56 &      h   &      h   &     0.15 &     0.12  \\
 Fe II 114  & 3468.68 &      h   &      h   &     0.30 &     0.25  \\
 Fe II 114  & 3493.47 &      h   &      h   &     0.28 &     0.18  \\
 Fe II] 16  & 3494.67 &      h   &      h   &     0.56 &     0.36  \\
 Fe II] 16  & 3507.40 &      h   &      h   &     0.15 &      n    \\
 Cr II] 2   & 3511.84 &      h   &      h   &     0.48 &     0.45  \\
 Ti II] 15  & 3552.85 &      h   &      h   &     0.02 &     0.06  \\
 Ti II] 15  & 3561.57 &      h   &      h   &     0.25 &     0.14  \\
 Ti II] 15  & 3573.74 &      h   &      h   &     0.55 &     0.46  \\
 Ti II] 15  & 3587.13 &      h   &      h   &     0.46 &     0.15  \\
 Ti II] 15  & 3596.05 &      h   &      h   &     0.11 &     0.07  \\
 Ti II] 76  & 3596.55 &      h   &      h   &     0.11 &     0.07  \\
 Ti II] 76  & 3608.89 &      h   &      h   &     0.33 &     0.16  \\
 Ti II] 76  & 3613.30 &      h   &      h   &     0.33 &     0.16  \\
 Ti II 52   & 3624.83 &      h   &      h   &     0.39 &     0.51  \\
 Cr II 12   & 3631.49 &      h   &      h   &     0.14 &      n    \\
 Cr II 12   & 3631.72 &      h   &      h   &     0.14 &      n    \\
 Cr II] 1   & 3651.68 &      h   &      h   &     0.21 &     0.06  \\
 Ti II 75   & 3659.76 &      h   &      h   &     0.00 &      n    \\
 Ti II 75   & 3662.24 &      h   &      h   &     0.04 &      n    \\
 Cr II 12   & 3677.69 &      h   &      h   &     0.06 &     0.06  \\
 Cr II 12   & 3677.86 &      h   &      h   &     0.06 &     0.06  \\
 Cr II 12   & 3677.93 &      h   &      h   &     0.06 &     0.06  \\
 Ti II 14   & 3685.19 &      h   &      h   &     0.15 &      n    \\
 Cr II 12   & 3712.97 &      h   &      h   &     0.14 &     0.08  \\
 Cr II 12   & 3713.04 &      h   &      h   &     0.07 &     0.04  \\
 Cr II 20   & 3715.19 &      h   &      h   &     0.08 &     0.05  \\
 Cr II] 117 & 3727.37 &      h   &      h   &     0.12 &     0.20  \\
 Cr II] 117 & 3737.55 &      h   &      h   &     0.03 &     0.05  \\
 Ti II 72   & 3741.63 &      h   &      h   &     0.22 &     0.20  \\
 Ti II 72   & 3757.68 &      h   &      h   &     0.07 &     0.05  \\
 Ti II 13   & 3759.29 &      h   &      h   &     0.15 &     0.10  \\
 Ti II 13   & 3761.32 &      h   &      h   &     0.15 &     0.10  \\
 Fe II] 29  & 3764.09 &      h   &      h   &     0.29 &     0.56  \\
 Ti II] 12  & 3813.39 &      h   &      h   &     0.03 &     0.05  \\
 Ti II] 12  & 3814.38 &      h   &      h   &     0.03 &     0.05  \\
 Fe II] 29  & 3824.91 &      h   &      h   &     0.30 &     0.21  \\
 Fe II] 23  & 3833.02 &      h   &      h   &     0.05 &     0.03  \\
 Mg II 5    & 3848.24 &      h   &      h   &     0.29 &     0.26  \\
 Mg II 5    & 3850.40 &      h   &      h   &     0.29 &     0.26  \\
 Fe II] 29  & 3872.77 &      h   &      h   &     0.16 &     0.17  \\
 Ti II 34   & 3900.55 &      h   &      h   &     0.05 &      n    \\
 Fe II] 29  & 3908.54 &      h   &      h   &     0.28 &     0.41  \\
 Ca II K    & 3933.66 &      h   &      h   &     0.22 &     0.28  \\
 Fe II] 3   & 3938.29 &      h   &      h   &     0.25 &     0.40  \\
 Fe II] 29  & 3964.57 &      h   &     0.04 &     0.06 &     0.09  \\
 Ca II H    & 3968.47 &      h   &     0.14 &     0.19 &     0.29  \\
 Fe II] 29  & 3974.16 &      h   &      n   &     0.11 &     0.08  \\
 Fe II] 29  & 4002.07 &      h   &      n   &     0.04 &     0.15  \\
 Ti II] 11  & 4012.37 &      h   &     0.16 &     0.14 &     0.11  \\
 Ti II] 11  & 4025.14 &      h   &     0.09 &     0.14 &     0.18  \\
 Ti II 87   & 4028.33 &      h   &     0.05 &     0.07 &     0.09  \\
 Cr II] 19  & 4051.97 &      h   &     0.01 &     0.04 &     0.03  \\
 Ti II 87   & 4053.81 &      h   &     0.01 &     0.04 &     0.03  \\
 
\hline
\end{tabular}
\end{center}
\end{table*}
\addtocounter{table}{-1}
\begin{table*}[ht]
\caption{(continued)} 
\begin{center}
\begin{tabular}{|l|c|c|c|c|c|}
\hline
Line & $\lambda (\AA)$ & KPNO 4-m & 2.3-m Bok & Keck & KPNO 2.1-m \\
\hline

 Cr II] 19  & 4054.11 &      h   &     0.01 &     0.04 &     0.03  \\
 Ti II 11   & 4056.21 &      h   &     0.01 &     0.04 &     0.03  \\
 Cr II 19   & 4063.94 &      h   &     0.22 &     0.13 &     0.36  \\
 Fe II] 21  & 4075.95 &      h   &     0.11 &     0.19 &     0.10  \\
 Fe II 28   & 4087.27 &      h   &     0.03 &     0.03 &     0.03  \\
 Fe II] 21  & 4119.52 &      h   &     0.07 &     0.08 &     0.07  \\
 Fe II 28   & 4122.64 &      h   &     0.11 &     0.13 &     0.12  \\
 Fe II 27   & 4128.73 &      h   &     0.19 &     0.32 &     0.24  \\
 Fe II] 12  & 4151.79 &      h   &     0.07 &     0.14 &      n    \\
 Ti II 21   & 4161.52 &      h   &     0.08 &     0.04 &     0.08  \\
 Ti II 105  & 4163.64 &      h   &     0.08 &     0.04 &     0.08  \\
 Ti II 105  & 4171.90 &      h   &     0.02 &     0.02 &     0.02  \\
 Fe II 27   & 4173.45 &      h   &     0.16 &     0.21 &     0.23  \\
 Ti II 105  & 4174.09 &      h   &     0.05 &     0.06 &     0.07  \\
 Fe II] 21  & 4177.69 &      h   &     0.21 &     0.28 &     0.30  \\
 Fe II 28   & 4178.86 &      h   &     0.20 &     0.26 &     0.28  \\
 Ti II 21   & 4184.33 &      h   &     0.09 &     0.06 &      n    \\
 Ti II 21   & 4190.29 &      h   &     0.03 &     0.02 &      n    \\
 Cr II 26   & 4207.35 &      h   &     0.27 &     0.28 &     0.28  \\
 Fe II 27   & 4233.17 &      h   &     0.69 &     0.68 &     0.78  \\
 Cr II 31   & 4242.38 &      h   &     0.32 &     0.43 &     0.46  \\
 Fe II 28   & 4258.16 &      h   &     0.18 &     0.15 &     0.18  \\
 Fe II] 21  & 4258.34 &      h   &     0.09 &     0.08 &     0.09  \\
 Cr II 31   & 4261.90 &      h   &     0.09 &     0.08 &     0.09  \\
 Fe II 27   & 4273.32 &      h   &     0.15 &     0.17 &     0.25  \\
 Cr II 31   & 4275.54 &      h   &     0.08 &     0.08 &     0.12  \\
 Fe II 32   & 4278.16 &      h   &     0.15 &     0.17 &     0.25  \\
 Cr II 31   & 4284.20 &      h   &     0.04 &     0.04 &     0.05  \\
 Ti II 20   & 4287.89 &      h   &     0.08 &     0.08 &     0.10  \\
 Ti II 41   & 4290.22 &      h   &     0.12 &     0.12 &     0.15  \\
 Ti II 20   & 4294.10 &      h   &     0.08 &     0.08 &     0.10  \\
 Fe II 28   & 4296.57 &      h   &     0.28 &     0.28 &     0.35  \\
 Ti II 41   & 4300.05 &      h   &     0.08 &     0.08 &     0.10  \\
 Ti II 41   & 4301.93 &      h   &     0.06 &     0.06 &     0.07  \\
 Fe II 27   & 4303.17 &      h   &     0.20 &     0.20 &     0.25  \\
 Ti II 41   & 4307.90 &      h   &     0.08 &     0.08 &     0.10  \\
 Ti II 41   & 4312.86 &      h   &     0.13 &     0.15 &     0.19  \\
 Fe II 32   & 4314.29 &      h   &     0.27 &     0.29 &     0.38  \\
 Ti II 41   & 4314.98 &      h   &     0.07 &     0.07 &     0.09  \\
 Ti II 94   & 4316.81 &      h   &     0.05 &     0.06 &     0.08  \\
 Ti II 94   & 4330.26 &      h   &     0.15 &     0.12 &     0.12  \\
 Ti II 41   & 4330.71 &      h   &     0.29 &     0.24 &     0.23  \\
 Ti II 20   & 4337.92 &      h   &     0.09 &     0.07 &     0.07  \\
 Fe II 32   & 4338.70 &      h   &     0.15 &     0.12 &     0.12  \\
 Ti II 20   & 4344.29 &      h   &     0.09 &     0.07 &     0.07  \\
 Ti II 94   & 4350.83 &      h   &     0.06 &     0.05 &     0.09  \\
 Fe II 27   & 4351.76 &      h   &     0.38 &     0.34 &     0.63  \\
 Fe II 148  & 4360.03 &      h   &     0.54 &     0.63 &     0.51  \\
 Ti II 104  & 4367.66 &      h   &     0.05 &     0.02 &     0.04  \\
 Fe II 28   & 4369.40 &      h   &     0.10 &     0.05 &     0.09  \\
 Fe II 33   & 4372.22 &      h   &     0.06 &     0.03 &     0.05  \\
 Ti II 93   & 4374.82 &      h   &     0.02 &     0.01 &     0.02  \\
 Fe II 32   & 4384.32 &      h   &     0.34 &     0.37 &     0.51  \\
 Fe II 27   & 4385.38 &      h   &     0.20 &     0.22 &     0.31  \\
 Ti II 104  & 4386.80 &      h   &     0.07 &     0.07 &     0.10  \\
 Ti II 61   & 4390.98 &      h   &     0.05 &     0.04 &     0.05  \\
 Ti II] 51  & 4394.07 &      h   &     0.08 &     0.07 &     0.08  \\
\hline
\end{tabular}
\end{center}
\end{table*}
\addtocounter{table}{-1}
\begin{table*}[ht]
\caption{(continued)} 
\begin{center}
\begin{tabular}{|l|c|c|c|c|c|}
\hline
Line & $\lambda (\AA)$ & KPNO 4-m & 2.3-m Bok & Keck & KPNO 2.1-m \\
\hline
 
 Ti II 19   & 4395.03 &      h   &     0.08 &     0.07 &     0.08  \\
 Ti II 61   & 4395.85 &      h   &     0.08 &     0.07 &     0.08  \\
 Ti II 61   & 4398.31 &      h   &     0.02 &     0.02 &     0.02  \\
 Ti II] 51  & 4399.79 &      h   &     0.16 &     0.15 &     0.16  \\
 Ti II 61   & 4409.22 &      h   &     0.03 &     0.03 &     0.04  \\
 Ti II 61   & 4409.48 &      h   &     0.03 &     0.03 &     0.04  \\
 Ti II 115  & 4411.08 &      h   &     0.06 &     0.07 &     0.09  \\
 Fe II 32   & 4413.60 &      h   &     0.16 &     0.16 &     0.21  \\
 Fe II 27   & 4416.82 &      h   &     0.32 &     0.33 &     0.43  \\
 Ti II] 40  & 4417.72 &      h   &     0.10 &     0.10 &     0.13  \\
 Ti II] 51  & 4418.34 &      h   &     0.16 &     0.16 &     0.21  \\
 Ti II 93   & 4421.95 &      h   &     0.10 &     0.10 &     0.13  \\
 Ti II] 40  & 4441.73 &      h   &     0.16 &     0.10 &     0.03  \\
 Ti II 19   & 4443.80 &      h   &     0.22 &     0.15 &     0.04  \\
 Ti II 31   & 4444.50 &      h   &     0.11 &     0.07 &     0.02  \\
 Ti II 19   & 4450.48 &      h   &     0.28 &     0.52 &     0.93  \\
 Fe II] 26  & 4461.43 &      h   &     0.07 &     0.04 &     0.03  \\
 Ti II] 40  & 4464.45 &      h   &     0.07 &     0.04 &     0.03  \\
 Ti II 31   & 4468.49 &      h   &     0.24 &     0.14 &     0.11  \\
 Ti II 18   & 4469.13 &      h   &     0.07 &     0.04 &     0.03  \\
 Ti II] 40  & 4470.86 &      h   &     0.07 &     0.04 &     0.03  \\
 Fe II 37   & 4472.92 &      h   &     0.36 &     0.61 &     0.59  \\
 Ti II 115  & 4488.32 &      h   &     0.07 &     0.06 &     0.10  \\
 Fe II 37   & 4489.18 &      h   &     0.29 &     0.26 &     0.38  \\
 Fe II 37   & 4491.40 &      h   &     0.29 &     0.26 &     0.38  \\
 Ti II] 18  & 4493.53 &      h   &     0.07 &     0.06 &     0.10  \\
 Ti II 31   & 4501.27 &      h   &     0.48 &     0.29 &     0.22  \\
 Fe II 38   & 4508.28 &      h   &     0.30 &     0.35 &     0.42  \\
 Fe II 37   & 4515.34 &      h   &     0.39 &     0.46 &     0.55  \\
 Fe II 37   & 4520.22 &      h   &     0.46 &     0.54 &     0.65  \\
 Fe II 38   & 4522.63 &      h   &     0.30 &     0.35 &     0.42  \\
 Ti II 82   & 4529.46 &      h   &     0.11 &     0.09 &     0.02  \\
 Ti II 50   & 4533.97 &      h   &     0.23 &     0.17 &     0.03  \\
 Fe II 37   & 4534.17 &      h   &     0.23 &     0.17 &     0.03  \\
 Fe II 38   & 4541.52 &      h   &     0.37 &     0.42 &     0.62  \\
 Fe II 38   & 4549.47 &      h   &     0.30 &     0.34 &     0.49  \\
 Cr II 44   & 4555.02 &      h   &     0.15 &     0.15 &     0.14  \\
 Fe II 37   & 4555.89 &      h   &     0.58 &     0.61 &     0.54  \\
 Cr II 44   & 4558.66 &      h   &     0.29 &     0.31 &     0.27  \\
 Ti II 50   & 4563.76 &      h   &     0.15 &     0.15 &     0.14  \\
 Ti II 82   & 4571.97 &      h   &     0.18 &     0.16 &     0.17  \\
 Fe II 38   & 4576.33 &      h   &     0.33 &     0.29 &     0.31  \\
 Fe II] 26  & 4580.06 &      h   &     0.18 &     0.16 &     0.17  \\
 Fe II 37   & 4582.83 &      h   &     0.37 &     0.33 &     0.34  \\
 Fe II 38   & 4583.83 &      h   &     0.31 &     0.28 &     0.29  \\
 Fe II] 26  & 4584.00 &      h   &     0.07 &     0.07 &     0.07  \\
 Cr II 44   & 4588.22 &      h   &     0.10 &     0.26 &     0.38  \\
 Cr II 44   & 4592.09 &      h   &     0.08 &     0.21 &     0.31  \\
 Fe II] 43  & 4601.38 &      h   &     0.61 &     0.36 &     0.05  \\
 Cr II 44   & 4616.61 &      h   &     0.04 &     0.05 &     0.08  \\
 Cr II 44   & 4618.84 &      h   &     0.07 &     0.10 &     0.16  \\
 Fe II 38   & 4620.51 &      h   &     0.18 &     0.25 &     0.39  \\
 Fe II 37   & 4629.34 &      h   &     1.09 &     1.04 &     1.03  \\
 Fe II 38   & 4648.23 &      h   &     0.08 &     0.06 &     0.03  \\
 Fe II] 25  & 4648.94 &      h   &     0.10 &     0.07 &     0.04  \\
 Fe II] 43  & 4656.98 &      h   &     0.14 &     0.11 &     0.06  \\
 Fe II] 44  & 4663.71 &      h   &     0.26 &     0.28 &     0.39  \\
 Fe II] 26  & 4665.80 &      h   &     0.04 &     0.04 &     0.06  \\
\hline
\end{tabular}
\end{center}
\end{table*}
\addtocounter{table}{-1}
\begin{table*}[ht]
\caption{(continued)} 
\begin{center}
\begin{tabular}{|l|c|c|c|c|c|}
\hline
Line & $\lambda (\AA)$ & KPNO 4-m & 2.3-m Bok & Keck & KPNO 2.1-m \\
\hline 

 Fe II 37   & 4666.75 &      h   &     0.13 &     0.14 &     0.19  \\
 Fe II] 25  & 4670.18 &      h   &     0.05 &     0.06 &     0.08  \\
 Fe II] 17  & 4724.07 &      h   &     0.21 &     0.10 &     0.46  \\
 Fe II] 31  & 4772.77 &      h   &     0.08 &     0.10 &     0.45  \\
 Ti II 17   & 4798.53 &      h   &     0.03 &     0.06 &     0.18  \\
 Ti II 92   & 4805.10 &      h   &     0.07 &     0.12 &     0.35  \\
 Fe II] 11  & 4818.26 &      h   &     0.35 &     0.30 &     0.61  \\
 Cr II 30   & 4836.22 &      h   &     0.35 &     0.53 &     0.89  \\
 Cr II 30   & 4848.24 &      h   &     0.25 &     0.17 &     0.30  \\
 Fe II] 25  & 4855.55 &      h   &     0.49 &     0.34 &     0.60  \\
 Fe II] 25  & 4871.28 &      h   &     0.13 &     0.13 &     0.27  \\
 Ti II 114  & 4874.03 &      h   &     0.13 &     0.13 &     0.27  \\
 Cr II 30   & 4876.41 &      h   &     0.10 &     0.09 &     0.20  \\
 Fe II] 36  & 4893.82 &      h   &     0.47 &     0.49 &     1.01  \\
 Ti II 114  & 4911.20 &      h   &     0.04 &     0.06 &     0.18  \\
 Fe II 42   & 4923.92 &      h   &     1.00 &     1.02 &     1.72  \\
 Fe II] 36  & 4924.92 &      h   &     0.20 &     0.20 &     0.34  \\
 Fe II] 36  & 4947.32 &      h   &     0.27 &     0.28 &     0.74  \\
 Ti II      & 4961.72 &     0.16 &     0.16 &     0.21 &     0.63  \\
 Fe II] 25  & 4991.13 &     0.21 &     0.19 &     0.20 &     0.29  \\
 Fe II] 36  & 4993.35 &     0.85 &     0.75 &     0.82 &     1.17  \\
 Fe II] 25  & 5000.73 &     0.09 &     0.08 &     0.08 &     0.12  \\
 Fe II 42   & 5018.45 &     1.73 &     1.29 &     1.36 &     1.76  \\
 Fe II] 36  & 5036.94 &     0.24 &     0.26 &     0.27 &     1.35  \\
 ?          & 5050.62 &     0.31 &     0.19 &     0.24 &     0.20  \\
 Ti II 113  & 5072.30 &     0.61 &     0.37 &     0.42 &     0.77  \\
 Fe II] 35  & 5100.66 &     0.64 &     0.36 &     0.38 &     0.58  \\
 Fe II] 35  & 5120.34 &     0.61 &     0.38 &     0.40 &     0.36  \\
 Fe II] 35  & 5132.67 &     0.29 &     0.20 &     0.24 &     0.18  \\
 Fe II] 35  & 5136.80 &     0.14 &     0.10 &     0.12 &     0.09  \\
 Fe II] 35  & 5146.13 &     0.34 &     0.24 &     0.29 &     0.22  \\
 Fe II] 35  & 5150.94 &     0.21 &     0.15 &     0.18 &     0.14  \\
 Fe II] 35  & 5154.40 &     0.17 &     0.12 &     0.15 &     0.11  \\
 Fe II] 35  & 5161.18 &     0.33 &     0.19 &      n   &     0.53  \\
 Fe II 42   & 5169.03 &     1.49 &     1.15 &     1.20 &     0.88  \\
 Fe II] 35  & 5171.74 &     0.37 &     0.29 &     0.30 &     0.22  \\
 Fe II 49   & 5197.57 &     1.87 &     1.36 &     1.32 &     1.38  \\
 Ti II 70   & 5226.53 &     0.40 &     0.44 &     0.41 &     0.33  \\
 Fe II 49   & 5234.62 &     1.56 &     0.96 &     0.97 &     1.26  \\
 Fe II 49   & 5254.92 &     0.37 &     0.45 &     0.38 &     0.25  \\
 Fe II 41   & 5256.89 &     0.13 &     0.16 &     0.13 &     0.09  \\
 Fe II 48   & 5264.80 &     0.83 &     0.51 &     0.51 &     0.77  \\
 Fe II 49   & 5275.99 &     0.83 &     0.51 &     0.51 &     0.77  \\
 Fe II 41   & 5284.09 &     1.00 &     0.97 &     0.85 &     0.54  \\
 Fe II 49   & 5316.61 &     0.59 &     0.45 &     0.41 &     0.45  \\
 Fe II 48   & 5316.78 &     1.48 &     1.13 &     1.04 &     1.13  \\
 Fe II 49   & 5325.56 &     0.15 &     0.11 &     0.10 &     0.11  \\
 Ti II 69   & 5336.81 &     0.06 &     0.05 &     0.04 &     0.03  \\
 Fe II 48   & 5337.71 &     0.28 &     0.27 &     0.19 &     0.16  \\
 Fe II 49   & 5346.56 &     0.41 &     0.27 &     0.38 &     0.16  \\
 Fe II 48   & 5362.86 &     0.86 &     0.65 &     0.47 &     0.72  \\
 Ti II 80   & 5367.95 &     0.26 &     0.20 &     0.14 &     0.22  \\
 Cr II] 29  & 5392.95 &     0.51 &     0.38 &     0.36 &     0.23  \\
 Fe II 48   & 5414.09 &     0.35 &     0.18 &     0.27 &     0.38  \\
 
\hline
\end{tabular}
\end{center}
\end{table*}
\addtocounter{table}{-1}
\begin{table*}[ht]
\caption{(continued)} 
\begin{center}
\begin{tabular}{|l|c|c|c|c|c|}
\hline
Line & $\lambda (\AA)$ & KPNO 4-m & 2.3-m Bok & Keck & KPNO 2.1-m \\
\hline

 Fe II 49   & 5425.27 &     0.82 &     0.68 &     0.55 &     0.55  \\
 Fe II] 55  & 5432.97 &     0.16 &     0.14 &     0.11 &     0.11  \\
 Ti II] 68  & 5446.46 &     0.22 &     0.04 &     0.15 &     0.06  \\
 Fe II 49   & 5477.67 &     0.17 &     0.13 &     0.13 &     0.07  \\
 Cr II 50   & 5478.35 &     0.08 &     0.06 &     0.06 &     0.03  \\
 Cr II 50   & 5502.05 &     0.14 &     0.11 &     0.12 &     0.08  \\
 Cr II 50   & 5503.18 &     0.14 &     0.11 &     0.12 &     0.08  \\
 Cr II 50   & 5508.60 &     0.14 &     0.11 &     0.12 &     0.08  \\
 Fe II] 56  & 5525.14 &     0.23 &     0.18 &     0.19 &     0.33  \\
 Ti II] 68  & 5529.94 &     0.11 &     0.09 &     0.09 &     0.17  \\
 Fe II] 55  & 5534.86 &     0.67 &     0.57 &     0.49 &     0.25  \\
 Fe II] 24  & 5864.54 &     0.08 &      n   &     0.20 &      n    \\
 Na I D     & 5889.89 &     0.43 &     0.75 &     0.32 &     0.54  \\
 Na I D     & 5895.92 &     0.57 &     0.25 &     0.56 &     0.21  \\
 Fe II] 47  & 5932.05 &     0.07 &     0.02 &     0.09 &     0.08  \\
 Fe II 182  & 5952.55 &     0.08 &      h   &     0.07 &     0.05  \\
 Fe II] 46  & 5991.39 &     0.22 &      h   &     0.33 &     0.22  \\
 Fe II] 46  & 6084.11 &     0.17 &     0.07 &     0.08 &     0.22  \\
 Fe II] 46  & 6113.33 &     0.10 &     0.10 &     0.06 &     0.14  \\
 Fe II] 46  & 6116.05 &     0.07 &     0.07 &     0.04 &     0.10  \\
 Fe II] 46  & 6129.71 &     0.05 &     0.05 &     0.03 &     0.07  \\
 Fe II 74   & 6147.74 &     0.41 &     0.32 &     0.28 &     0.37  \\
 Fe II 74   & 6149.25 &     0.41 &     0.32 &     0.28 &     0.37  \\
 Fe II] 46  & 6196.71 &     0.21 &     0.19 &     0.21 &     0.22  \\
 Fe II 74   & 6238.37 &     0.25 &     0.43 &     0.22 &     0.07  \\
 Fe II] 34  & 6239.37 &     0.07 &     0.13 &     0.07 &     0.02  \\
 Fe II 74   & 6239.95 &     0.10 &     0.17 &     0.09 &     0.03  \\
 Fe II 74   & 6247.55 &     0.78 &     0.22 &     0.76 &     1.27  \\
 Fe II] 34  & 6279.83 &     0.16 &     0.10 &     0.23 &     0.18  \\
 Fe II] 34  & 6307.53 &     0.16 &     0.10 &     0.23 &     0.18  \\
 ?          & 6338.80 &     0.34 &     0.21 &     0.36 &     0.24  \\
 Fe II 40   & 6369.45 &     0.58 &     0.51 &     0.55 &     0.66  \\
 Fe II 74   & 6407.30 &     0.23 &     0.17 &     0.17 &     0.17  \\
 Fe II 74   & 6416.89 &     0.47 &     0.34 &     0.34 &     0.35  \\
 Fe II 40   & 6432.68 &     0.31 &     0.27 &     0.15 &     0.38  \\
 Fe II 74   & 6456.39 &     0.91 &     0.68 &     0.47 &     0.58  \\
 Fe II 40   & 6516.05 &     0.76 &     0.85 &     0.54 &     0.83  \\
\noalign{\smallskip}
\hline
\end{tabular}
\end{center}
\end{table*}


 \end{document}